\documentclass[lettersize,journal]{IEEEtran}
\usepackage{amsmath,amsfonts}
\usepackage{amsthm}
\usepackage{algorithmic}
\usepackage{algorithm}
\usepackage{array}
\usepackage{bbm}
\usepackage{textcomp}
\usepackage{stfloats}
\usepackage{url}
\usepackage{verbatim}
\usepackage{enumitem}
\usepackage{graphicx}
\usepackage{lipsum}
\usepackage{multicol}
\usepackage{float}
\usepackage{bbding}
\usepackage{subcaption}
\usepackage{hyperref}
\usepackage{cleveref}

\usepackage{svg}

\crefname{figure}{fig}{figures}
\Crefname{figure}{Fig}{Figures}
\usepackage{cite}
\newtheorem{definition}{Definition}
\newtheorem{theorem}{Theorem}

\begin{document}

\title{Attribute-Based Encryption With Payable Outsourced Decryption Using Blockchain and Responsive Zero Knowledge Proof}

\author{Dongliang~Cai, Borui~Chen, Liang~Zhang, Kexin~Li
        and~Haibin~Kan
\thanks{Dongliang Cai, Borui Chen and  Kexin Li are with the Shanghai Key Laboratory of Intelligent Information Processing, School of Computer Science, Fudan University, Shanghai 200433, China, and also with the Shanghai Engineering Research Center of Blockchain, Shanghai 200433, China (e-mail:
22110240060@m.fudan.edu.cn; 23210240072@m.fudan.edu.cn; 24110240170@m.fudan.edu.cn).}
\thanks{Liang Zhang is with Department of Industrial Engineering and Decision Analytics, Hong Kong University of Science and Technology. (e-mail: scottzhang@ust.hk)}
\thanks{Haibin Kan is with the Shanghai Key Laboratory of Intelligent Information
Processing, School of Computer Science, Fudan University, Shanghai 200433,
China, also with the Shanghai Engineering Research Center of Blockchain,
Shanghai 200433, China, and also with the Yiwu Research Institute of Fudan
University, Yiwu 322000, China (e-mail: hbkan@fudan.edu.cn)}
}



\maketitle

\begin{abstract}
Attribute-Based Encryption (ABE) is a promising solution for access control in cloud services. However, the heavy decryption overhead hinders its widespread adoption. A general approach to address this issue is to outsource decryption to decryption cloud service(DCS). Existing schemes have utilized various methods to enable users to verify outsourced results; however, they lack an effective mechanism to achieve exemptibility which enables the honest DCS to escape from wrong claims. And it is impractical to assume that the DCS will provide free services. In this paper, we propose a blockchain-based payable outsourced decryption ABE scheme that achieves both verifiability and exemptibility without adding redundant information to ABE ciphertext. We use zero-knowledge proof to verify outsourced results on blockchain and introduce an optional single-round challenge game under optimistic assumption to address the high cost of proof generation. Moreover, our system achieves fairness and decentralized outsourcing to protect the interests of all parties. Finally, we implement and evaluate our scheme on Ethereum to demonstrate its feasibility and efficiency, the gas usage in attribute numbers from 5 to 60 is 11$\times$ to 140$\times$ in the happy case and  4$\times$ to 55$\times$ in the challenge case lower than the scheme of Ge et al. (TDSC'23).
\end{abstract}


\section{Introduction}
\IEEEPARstart{W}{ith} the rapid development of cloud computing, cloud storage services have fundamentally reshaped data sharing. Increasingly, users and organizations are uploading and sharing data via cloud storage services. However, since Cloud Service Providers (CSPs) are not entirely trustworthy and much of the data stored in the cloud is highly sensitive, such as personal medical records and internal corporate data, access control in cloud storage becomes a critical issue. In public cloud storage, access control can be achieved through functional encryption \cite{boneh2011functional}, among which Ciphertext-Policy Attribute-Based Encryption (CP-ABE) is considered one of the most suitable solutions \cite{xue2019attribute}. CP-ABE enables fine-grained access control by embedding access control policy into the ciphertext and attributes into the keys. Data owners can design access policy at the encryption phase, and decryption is possible only if the set of data user's attributes satisfies the access policy embedded in the ciphertext.

However, one major issue in functional encryption is the heavy decryption overhead. For example, in pairing-based attribute-based encryption, the pairing operations required during decryption are typically linear to the size of the access control policy \cite{goyal2006attribute},\cite{bethencourt2007ciphertext},\cite{waters2011ciphertext}, which is particularly inefficient for lightweight and mobile devices. To address this issue, Green et al. \cite{green2011outsourcing} proposed an outsourced decryption scheme for attribute-based encryption (OABE), which significantly reduces the computational overhead for users without revealing sensitive information about the original data. In this scheme, the user generates a transformation key (TK) and a retrieve key (RK) from the attribute private key (SK) and sends the TK and ciphertext (CT) to the decryption cloud server (DCS) for partial decryption, which transforms the ciphertext into an ElGamal ciphertext \cite{elgamal1985public}. The user then uses the RK to decrypt the ElGamal ciphertext and retrieve the plaintext.

Since the DCS may not perform the computation honestly to save computation resources, many verifiable outsourced decryption schemes for attribute-based encryption have been proposed in the literature \cite{lai2013attribute},\cite{li2013securely},\cite{qin2015attribute},\cite{lin2015revisiting},\cite{miao2023verifiable}. These schemes add redundant information to both the original ABE ciphertext and the transformed ciphertext to verify the correctness of the outsourced computation result. However, this redundant information increases the computational cost for the DCS during the transformation phase, potentially raising the user’s expenses. Furthermore, many previous schemes fail to achieve the properties of exemptibility and fairness. The exemptibility property ensures that the DCS is not falsely accused of malicious behavior if it returns a correct transformed ciphertext, while the fairness property ensures that the DCS gets paid if and only if it returns a correct result. Intuitively, these two properties could be achieved by introducing a trusted intermediary. Cui et al. \cite{cui2020pay} combined decentralized blockchain technology\cite{nakamoto2008bitcoin} to propose a functional encryption with payable outsourced decryption (FEPOD) scheme, which achieves these two properties without relying on a trusted authority. However, this scheme still introduces redundant information. Ge et al. \cite{ge2023attribute} eliminates redundant information and achieves decentralized outsourcing through delegating computation to the blockchain, but outsources heavy computation to smart contracts, which will cause high gas fees and performance bottlenecks. A straightforward method to reduce on-chain computation is zero-knowledge proof\cite{goldwasser1989knowledge}, but will lead to large computation with the proof generation.

\subsection{Motivation and Contribution}
Although many existing attribute-based encryption (ABE) schemes with outsourced decryption can detect malicious behavior by the decryption cloud server, they rely on additional redundant information in transformation, which increases the honest server's computational cost. Furthermore, schemes that achieve exemptibility and fairness without trust authority rely on smart contracts for heavy computation, leading to high gas fees and performance issues. Zero-knowledge proof is a general method for reducing on-chain computation; however, they also incur substantial overhead in off-chain proof generation. In this paper, we aim to design an attribute-based encryption scheme with payable outsourced decryption that overcomes the above issues, achieves desirable properties, and minimizes overhead. In summary, the contributions of this paper are as follows.
\begin{itemize}
    \item We propose a zk-friendly payable outsourced decryption attribute-based encryption (ABE) scheme without redundant information based on blockchain, which achieves verifiability, exemptibility, fairness and decentralized outsourcing, and the payment can be processed by blockchain cryptocurrency.
    \item We use zero-knowledge proof (ZKP) to verify outsourced decryption result on blockchain with constant gas fee, introduce a single-round challenge mechanism to greatly reduce proof generation with high computational cost through responsive ZKP, and achieve self-challenge attack resistance.
    \item We implement a concrete payable outsourced decryption ABE scheme with responsive ZKP on native Ethereum, develop the ZKP circuit using the widely adopted framework Halo2\cite{halo2} and evaluate feasibility and performance on Ethereum.
\end{itemize}

\subsection{Related Work}
\begin{itemize}[itemsep=1em]
    \item \textbf{Outsource Decryption Attribute Based Encryption.}

    \hspace{1em}Attribute-Based Encryption was first introduced as fuzzy identity-based encryption by Sahai and Waters in \cite{sahai2005fuzzy}. ABE schemes can be divided into two categories: Ciphertext-Policy ABE (CP-ABE) and Key-Policy ABE (KP-ABE) \cite{goyal2006attribute}, depending on the access policy is embedded into the ciphertext or the user's private key. A user can decrypt a ciphertext only if the set of attributes satisfies the access policy. However, the decryption involved in the ABE is usually too expensive, and this problem is especially acute for resource limited devices such as mobile devices, which greatly hinders its practical popularity. In order to reduce the decryption overhead for a user to recover the plaintext, Green et al. \cite{green2011outsourcing} suggested to outsource the majority of the decryption work without revealing the plaintext.

    \hspace{1em}However, \cite{green2011outsourcing} lacks a mechanism to verify whether the decryption cloud server has returned a correct transformed ciphertext. Consequently, Lai et al. \cite{lai2013attribute} modified the original model of ABE with outsourced decryption to include verifiability, but doubled the size of the underlying ABE ciphertext and the computation costs. It appended a redundant ciphertext of a random message and a tag to each ciphertext, and required the original untransformed ciphertext as an auxiliary input in the final decryption step by the user. Based on the scheme \cite{lai2013attribute}, Qin et al. \cite{qin2015attribute} and Lin et al. \cite{lin2015revisiting} optimized the efficiency of OABE. \cite{qin2015attribute} halved the ciphertext size in both server and client and is 2 to 4 times faster in decryption compared with the scheme \cite{lai2013attribute} and proposed an approach to convert any ABE scheme with outsourced decryption into an ABE scheme with verifiable outsourced decryption. \cite{lin2015revisiting} proposed a more efficient and generic construction of ABE with verifiable outsourced decryption based on an attribute-based key encapsulation mechanism, a symmetric-key encryption scheme, and a commitment scheme. Moreover, Li et al. \cite{li2013securely} proposed the notion of attribute-based encryption with both outsourced encryption and decryption, and considered to offload the overhead computation at authority by outsourcing key-issuing. However, \cite{li2013securely} only supports the threshold access policy. Following their work, Ma et al. \cite{ma2015verifiable} proposed an attribute-based encryption with both outsourced encryption and decryption that supports any monotonic access policy and defined stronger verifiability than \cite{lai2013attribute}. It claimed that even a user who has the private key cannot accuse the decryption cloud server of outputting incorrect results while it was not the case. But in fact they can not achieve the exemptibility since the user may reveal an incorrect private key and claim that the decryption cloud server has returned incorrect transformed ciphertext. Moreover, Wang et al. \cite{wang2019fully} proposed fully accountable ABE for the pay-as-you-go model, which can achieve high decryption efficiency for data consumers and prevention of Distributed Denial of Services(DDoS) attacks on the cloud ciphertexts. 

    \hspace{1em}None of the above schemes can achieve both exemptibility and fairness. To address this issue without central authority, Cui et al. \cite{cui2020pay} proposed an attribute-based encryption scheme using blockchain that outsources decryption to a smart contract. In their scheme, the user creates a smart contract that contains the decryption task. Any miner on the blockchain can initiate a transaction that calls the smart contract and receive the reward. However, this scheme still introduces redundant information and outsources heavy computation to smart contracts. Ge et al. \cite{ge2023attribute} eliminated redundant information and mitigated heavy computation problem of smart contracts through decomposing to pieces of meta computations, but still caused high gas usage and performance bottlenecks. Moreover, since native Ethereum\cite{wood2014ethereum} does not support pairing computation operation, this scheme is not compatible with native Ethereum. Recently, Tao et al. \cite{tao2024orr} tackled the issue of significant computational burden on outsourcing devices whose resources are not infinite in practice. They proposed a secure and efficient outsourced ABE scheme with decryption results reuse. Moreover, Mahdavi et al. \cite{mahdavi2024iot} presented novel precomputed and short ciphertext precomputed versions of a special case of ABE which are tailored for resource-constrained low-end IoT devices. In this scheme, some outsourcing methods for heavy computational operations are suggested to be used, and the data user can ensure verifiability by repeatedly querying and comparing the consistency of the results. Hou et al. \cite{hou2024blockchain}
     proposed a blockchain-based efficient verifiable outsourced attribute-based encryption in cloud, utilizing the technique of batch verification to make data users verify the decrypted plaintexts. In this scheme, the audit of verification results from the data user is performed by miners. 

    \item \textbf{Applications of Fraud Proof and Validity Proof on Blockchain.}

    \hspace{1em}The fraud proof assumes optimistic condition, defaulting to accept the statement unless disproven. It includes a challenge window during which challengers may initiate disputes to demonstrate the incorrectness of the statement. The validity proof provides evidence that a statement is correct, and acceptance occurs only when correctness is definitively established under a pessimistic assumption.

    \hspace{1em}Optimistic rollup\cite{bousfield2022arbitrum},\cite{optimism},\cite{teutsch2024scalable},\cite{armstrong2021ethereum},\cite{morph},\cite{Fuel} as one of the important solutions for Ethereum's layer2, rely on the optimistic assumption that most transactions will be valid, which enables more rapid and efficient transaction processing. To safeguard against invalid transactions, fraud proofs are employed to challenge and dispute them. Many projects\cite{bousfield2022arbitrum},\cite{optimism} employ the interactive bisection protocol\cite{kalodner2018arbitrum} to implement multi-round fraud proof, while others\cite{morph},\cite{Fuel} utilize zero-knowledge proof to achieve single-round fraud proof and shorten the challenge window. This fraud proof mechanism has been widely adopted across various blockchain applications beyond optimistic rollups, including the Lightning Network\cite{poon2016bitcoin}, Plasma\cite{poon2017plasma}, and distributed key generation protocol\cite{schindler2019ethdkg}.

     \hspace{1em}Zero-knowledge proof is an important cryptographic tool for another Ethereum layer2 solution, zero-knowledge Rollup, which uses validity proof instead of fraud proof. There are now many projects\cite{starkware}\cite{scroll}\cite{taiko} that leverage zero-knowledge proof technology to implement ZK-EVM(Ethereum Virtual Machine) or ZK-VM(Virtual Machine). These projects generate a proof for a batch of transactions on Layer 2 and submit it to Layer 1. Instead of re-executing all transactions, nodes can verify the transactions and smart contracts by checking the proof and updating the state. This approach significantly enhances the transaction throughput of the blockchain.
\end{itemize}

\subsection{Organization}
The rest of this paper is organized as follows. In \Cref{sec:Preliminaries}, we briefly revisit the definitions that are relevant to this paper. In \Cref{sec:System Overview}, we describe the system overview of our scheme. In \Cref{sec:Construction}, we present a generic construction of POABE scheme using blockchain and responsive zero-knowledge proof. In \Cref{sec:Performance And Evaluation}, we implement the proposed concrete scheme on Ethereum to evaluate its performance. Finally, this paper is concluded in \Cref{sec:Conclusion}.

\section{Preliminaries}
\label{sec:Preliminaries}
In this section, we review some definitions associated with our scheme.
\subsection{Bilinear Map}
Let $\mathbb{G}, \mathbb{G}_T$ are two cyclic groups of prime order $p$, $e$ is the bilinear map $\mathbb{G} \times \mathbb{G} \rightarrow \mathbb{G}_T$ and $g$ is the generator of $\mathbb{G}$. The bilinear map has three properties:
\begin{itemize}
    \item \textit{Bilinearity}: $e(g^a,g^b) = e(g,g)^{ab}$ holds for all $a,b \in \mathbb{Z}_{p}^{*}$, where $\mathbb{Z}_{p}$ is a field of prime order $p$.
    \item \textit{Non-degeneracy}: $e(g,g) \neq 1$.
    \item \textit{Computability}: $e(g,g)$ can be computed in polynomial time.
\end{itemize}
\subsection{Ciphertext-Policy Attribute-Based Encryption}
A ciphertext-policy attribute based encryption scheme \cite{bethencourt2007ciphertext} consists of four algorithms:
\begin{itemize}
    \item $Setup(\lambda) \rightarrow PK,MSK$. The Setup algorithm takes as input the security parameter and outputs the public key PK and a master key MSK.
    \item $KeyGen(MSK,S) \rightarrow SK$.  The KeyGen algorithm takes as input the master key MSK and a set of attributes S. It outputs a secret key SK.
    \item $Encrypt(M, PK, \Gamma) \rightarrow CT$. The Encrypt algorithm takes as input a message M, the public key PK and an access structure $\Gamma$. It outputs ciphertext CT that can only be decrypted by users who satisfy the specified access structure $\Gamma$.
    \item $Decrypt(PK,CT,SK) \rightarrow M$. The Decrypt algorithm takes as input the public key PK, a ciphertext CT and a secret key SK which is associated with a set S of attributes. If the attribute set S satisfies the access structure, this algorithm will successfully decrypt the ciphertext and return a message M.
\end{itemize}

\subsection{Blockchain and Smart Contracts}
Ethereum is a decentralized and open-source blockchain platform that incorporates smart contract functionality\cite{wood2014ethereum}. It enables the development and execution of decentralized applications(DApps) using smart contracts, which are self-executing contracts with the terms of the agreement directly written into code. Ethereum's Turing-complete Ethereum Virtual Machine (EVM) facilitates the execution of smart contract code, providing a flexible and programmable environment.

At the core of Ethereum’s structure are accounts, which are fundamental components of its system. Ethereum employs two types of accounts: externally owned accounts (EOAs), also referred to as user accounts, and smart contract accounts. Both types of accounts can hold and transfer the platform's native cryptocurrency, Ether (ETH), to other accounts, initiate new smart contracts, or invoke public functions of existing contracts. Additionally, each account is identified by a unique address within the blockchain and state, enabling interactions and transactions across the network.

This integration of blockchain technology with smart contracts allows Ethereum to go beyond simple transactions, facilitating the creation of complex, decentralized applications that can operate without intermediaries.

\subsection{Zero-Knowledge Proof}
Zero-knowledge proof (ZKP) was proposed by S. Goldwasser, S. Micali, and C. Rackoff in \cite{goldwasser1989knowledge}. Succinct non-interactive arguments of knowledge (SNARKs) \cite{kilian1992note},\cite{micali1994cs} are powerful cryptographic tools for checking certain statements with succinct proofs and fast verifications, leading to their widespread application in blockchain.

\begin{definition}
(\textbf{Non-Interactive Proof System}). A pair of probabilistic machines (P, V) is called a non-interactive proof system for a language L if V is polynomial-time and the following two conditions hold:
\begin{itemize}
    \item \textit{Completeness}: For every $x \in L$,
    $$
    Pr[V(x, R, P(X,R) = 1)] \geq 1 - negl(|x|)
    $$
    \item \textit{Soundness}: For every $x \notin L$ and every algorithm $B$,
    $$
    Pr[V(x, R, B(X,R) = 1)] \leq negl(|x|)
    $$
\end{itemize}
where $negl(|x|)$ is a negligible function of $|x|$ and R is a random variable uniformly distributed in $ \{0,1\}^{poly(|x|)}$, which is called the common reference string(CRS).
\end{definition}

\begin{definition}
(\textbf{Indexed Relation}). An \textbf{indexed relation} $\mathcal{R}$ is a set of triples $(\mathbbm{i}, \mathbbm{x}, \mathbbm{w})$ where $\mathbbm{i}$ is the index, $\mathbbm{x}$ is the public input, $\mathbbm{w}$ is the witness. For example, the indexed boolean circuit satisfiability relation $\mathcal{R}$ consists of triples $(\mathbbm{i}, \mathbbm{x}, \mathbbm{w})$, where $\mathbbm{i}$ is the description of a boolean circuit, $\mathbbm{x}$ is the partial assignment of the input and $\mathbbm{w}$ is the remaining assignment such that the output of the circuit is $1$. Let $\mathcal{L}(\mathcal{R})$ be the corresponding language, i.e., $\mathcal{L} (\mathcal{R}) = \left\{ (\mathbbm{i}, \mathbbm{x}): \exists \mathbbm{w}, \text{ such that } (\mathbbm{i}, \mathbbm{x}, \mathbbm{w}) \in \mathcal{R} \right\}$. 
\end{definition}

\begin{definition}
(\textbf{Preprocessing zk-SNARK}). A \textbf{preprocessing zk-SNARK} for a NP relation $\mathcal{R}$ is a triple of probabilistic polynomial-time(PPT) algorithms(Preprocess, Prove, Verify) defined as follows.
\begin{itemize}
    \item $Preprocess(\lambda, \mathcal{C})\rightarrow (pk,vk)$ On input a security parameter $\lambda$ and an arithmetic circuit $\mathcal{C}$ of size polynomial in $\lambda$ represents a NP relation $\mathcal{R}$, it outputs proving key $pk$ and verifying key $vk$.
    \item $Prove(pk,\mathbbm{x},\mathbbm{w})\rightarrow \pi$. On input proving key $pk$, a statement $\mathbbm{x}$ and a witness $\mathbbm{w}$, it generates a proof $\pi$.
    \item $Verify(vk,\mathbbm{x},\pi)\rightarrow accept/reject$. On input verifying key $vk$, a statement $\mathbbm{x}$ and a proof $\pi$, it outputs accept to indicate a valid proof and reject otherwise.
\end{itemize}
\end{definition}

\section{System Overview}
\label{sec:System Overview}
In this section, we describe our system architecture, threat model and system requirements.

\begin{figure}[ht]
    \centering
    \includegraphics[width=0.5\textwidth]{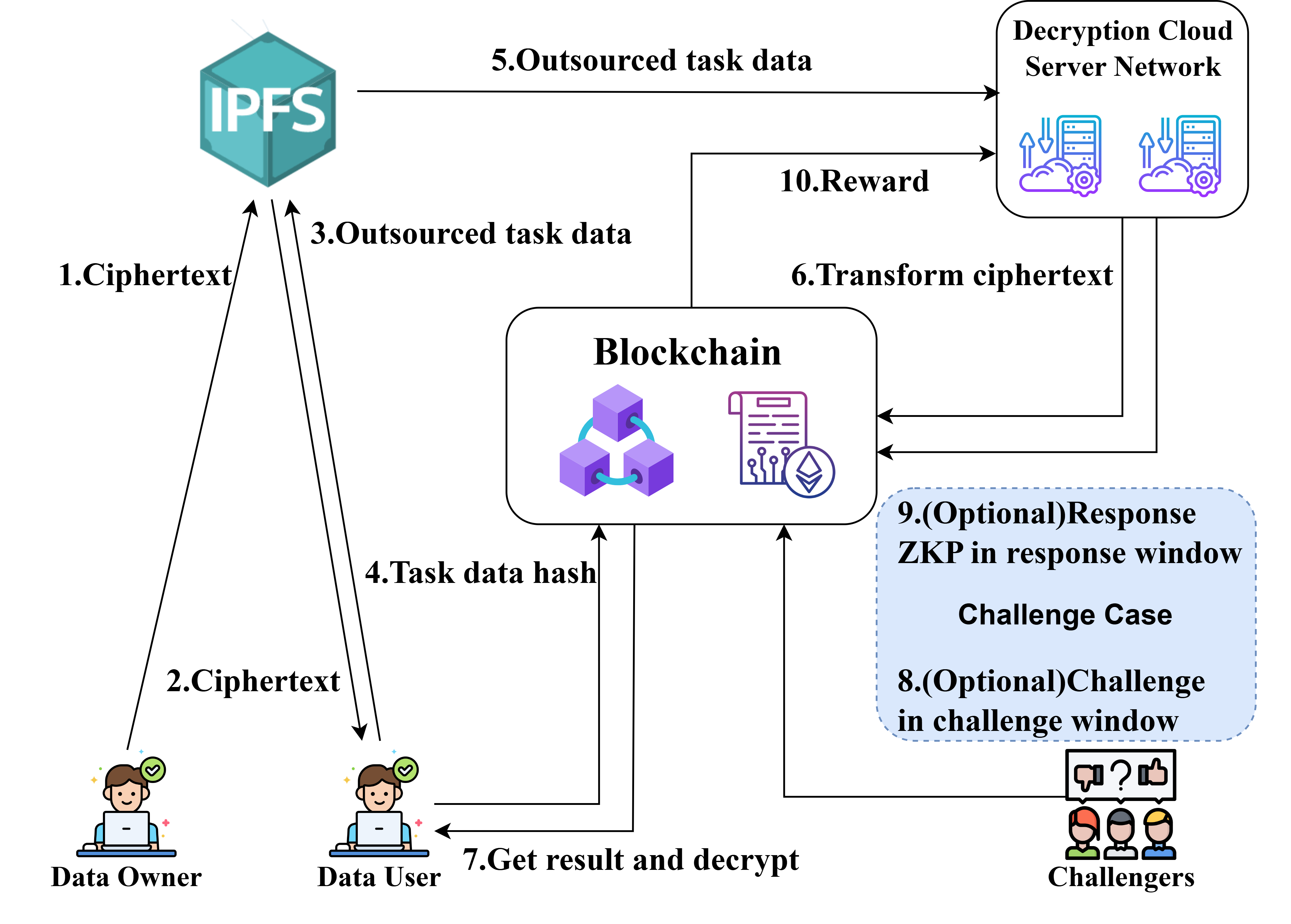}
    \caption{Overview of our scheme.}
    \label{fig:system_overview}
\end{figure}

\subsection{System Architecture}
 Our system architecture consists of five entities, ie.  Data Owners(DOs), InterPlanetary File System(IPFS), Data Users(DUs), Decryption Cloud Server(DCS) Network and Challengers. The role of each entity is defined as follows:
\begin{itemize}
    \item DOs uses a specified access structure to encrypt data and then upload the ciphertext to IPFS.
    \item IPFS is an open and decentralized system to store data without a central server, which stores ciphertexts and provides data according to the user's requirements in our system.
    \item DUs get ciphertext from IPFS, and then publish outsourcing decryption request on blockchain and original decryption task data to IPFS, once the correct partial decryption results are obtained, the final decryption can be accomplished.
    \item DCS network consists of many servers that compete for tasks, they stake tokens to blockchain at first to be eligible to submit result, execute decryption task, and submit result to blockchain to get reward. If the result is challenged within the challenge window, they need to send a valid ZK proof for corresponding computation within the response window.
    \item Challengers verify the result submitted by DCSs and they can stake tokens to challenge within the challenge window. Anyone, including DU, can be a challenger, DU can check the correctness of the results based on subsequent decryption while others can verify through re-computation.
\end{itemize}

\Cref{fig:system_overview} shows the overview of the happy and optional challenge case of our scheme, where the challenge case has two more steps of challenge and response than the happy case. The happy case consists of the following steps:
\begin{enumerate}
    \item DO encrypts a message using a symmetric key initially, followed by encrypting the symmetric key with a specified access structure before uploading ciphertexts to IPFS.
    \item DU obtains the ciphertexts from IPFS.
    \item DU generates the data required for outsourcing, and then sends the outsourcing data to IPFS.
    \item DU sends the hash of the outsourcing data to the smart contract along with reward.
    \item DCS observes decryption request on-chain and gets outsourcing data from IPFS.
    \item DCS executes the computation task and submits results to the smart contract.
    \item DU obtains the transform ciphertext from blockchain and decrypts it locally.
    \item Following the challenge window with no challenge, DCS receives the reward for the outsourced decryption.
\end{enumerate}

In the case of challenge, two additional steps will be involved:
\begin{enumerate}
    \item The challenger stakes tokens to the smart contract to challenge within the challenge window.
    \item DCS must submit a proof that can pass the smart contract verification within the response window. Otherwise, DCS fails and loses the staked tokens, while the challenger wins and receives the reward, and DU receives compensation. Conversely, if DCS wins, meaning the computation result is correct, it receives the tokens staked by the challenger as the cost of generating the proof.
\end{enumerate}

Note that in our encryption, we employ a Key Encapsulation Mechanism (KEM), encrypt the plaintext using symmetric encryption, and perform attribute-based encryption on the symmetric key, which not only reduces the computational overhead for DO during CP-ABE encryption but also prompts DU to challenge incorrect results computed by DCS.

\subsection{Threat Model}
In our scheme, DOs, blockchain, and IPFS are fully trusted. We consider that DCSs are malicious, which may perform incorrect outsourced decryption or even return a random value to save computation resources. Moreover, they may challenge themselves to attack our system. Due to collusion attacks and the potential for malicious accusations that DCS has returned incorrect results to avoid payment, we also assume that DUs are malicious.

\subsection{System Requirements}
\begin{itemize}
    \item Data Confidentiality: The data confidentiality property indicates that the unauthorized DUs cannot decrypt the transformed ciphertexts. In this work, this property is guaranteed by the first OABE scheme\cite{green2011outsourcing} and we adopted the security notion of against chosen plaintext attack (CPA) secure which is described in \cite{green2011outsourcing}.
    \item Verifiability: The verifiability means that malicious behavior of the decryption cloud server will be detected. Many previous works\cite{lai2013attribute},\cite{li2013securely},\cite{qin2015attribute},\cite{lin2015revisiting},\cite{miao2023verifiable} resort to adding redundant information to the ABE ciphertext and the transformed ciphertext to achieve the verifiability. And scheme in \cite{ge2023attribute} removes redundant information by claiming that the correctness of the transformed ciphertext is guaranteed by the Ethereum consensus protocol. To eliminate redundant information and reduce the costly overhead associated with Ethereum consensus computation, we utilize zero-knowledge proofs and on-chain smart contracts for verification in the challenge phase. And all data related to the outsourced decryption process is publicly accessible, we implement a challenge mechanism with a token incentive to motivate public verifiers to conduct off-chain verification and initiate on-chain challenges.
    \item Exemptibility: The exemptibility ensures that DCS is not falsely accused. \cite{ge2023attribute} depends on the consensus protocol while we utilize zero-knowledge proofs and challenge mechanism for DCS to demonstrate the correctness of previously submitted results during the challenge phase to achieve this property.
    \item Fairness: The fairness property in our work guarantees that: 
    \begin{itemize}
        \item DCS gets paid if it has returned a correct result and otherwise will be slashed.
        \item DU must pay for the correct result.
        \item Challenger wins the challenge will get reward and loses staked tokens otherwise.
    \end{itemize}
    The challenge mechanism with blockchain token incentive achieves fairness property, which incentivizes each party to perform the protocol honestly.
    \item Self-challenge attack resistance: With the introduction of the challenge mechanism, we need to prevent the DCS that submit the results from self-challenge which may makes delay and adversely affects the interests of DU. We need to design a mechanism that render it economically unfeasible for malicious DCS to cause delays through self-challenge.
\end{itemize}

\section{Construction}
\label{sec:Construction}
In this section, we present our construction of CP-ABE with payable outsourced decryption(CP-POABE) scheme based on Ethereum and responsive zero-knowledge proof. We will first present our zk-friendly CP-POABE scheme based on Green et al.’s OABE scheme \cite{green2011outsourcing}, and then describe a single-round challenge game and how to achieve constant gas consumption on Ethereum by zero-knowledge proof. Finally, we present the concrete smart contract design for our payable outsourced decryption ABE scheme.
\subsection{Proposed CP-ABE With Outsourced Decryption Scheme}
Our proposed CP-ABE with outsourced decryption scheme is as follows.
\begin{itemize}
    \item $Setup(\lambda, U) \rightarrow PK,MSK$. The setup algorithm takes as input a security parameter $\lambda$ and a universe description U. The public key generator (PKG) generates the public bilinear pairing parameters $BP = (\mathbb{G}, \mathbb{G}_T, e, p, g)$ and then choose a hash function $F:\{0,1\}^* \rightarrow \mathbb{G}$ that maps an attribute to group element. In addition, it chooses random elements $a, \alpha \in \mathbb{Z}_p$. Finally, it sets the master key MSK and public parameters PK as:
    $$
    MSK = (g^{\alpha}), PK = (g, e(g,g)^{\alpha}, g^a, F) 
    $$
    \item $KeyGen(PK, MSK, S) \rightarrow SK$. The keygen algorithm takes as input the public parameters PK, the master key MSK and a attribute set S. It chooses a random value $r \in \mathbb{Z}_{p}^{*}$ and compute:
    $$
    K = g^{\alpha}g^{ar},L =  g^r,\{K_{x} = F(x)^r\}_{x\in S}
    $$
    It sets the private key SK as:
    $$
    SK = (S,K,L,\{K_{x}\}_{x \in S})
    $$
    \item $Encrypt(PK, \mathcal{M}, (M,\rho)) \rightarrow CT$. The encrypt algorithm takes as input the public parameters PK and a message $\mathcal{M}$(a symmetric key in KEM setting) to encrypt. In addition, it takes as input an LSSS access structure $(M,\rho)$ where M is an $l \times n$ matrix and $\rho$ maps each row of $M$ to an attribute. The algorithm first selects a random vector $\vec{v} = (s,y_2,...,y_n) \in \mathbb{Z}_{p}^{n}$. For $i = 1$ to $l$, it computes $\lambda_i = \vec{v} \cdot M_i$, where $M_i$ is the ith row of $M$. In addition, it chooses random value $t_1,...,t_l \in \mathbb{Z}_{p}$ and compute:
    $$
    C = \mathcal{M} \cdot e(g,g)^{\alpha s},C' = g^s
    $$
    $$
    C_i = g^{a\lambda_i} \cdot F(\rho(i))^{-t_i}, D_i = g^{t_i}, 1 \leq i \leq l 
    $$
    It sets the ciphertext CT as:
    $$
    CT = ((M,\rho), C, C', \{C_i, D_i\}_{i\in {[1,...,l]}})
    $$
    \item $TKGen_{out}(SK) \rightarrow TK, RK$. The transform key generation algorithm is executed by DO and takes as input the private key SK and choose a random value $z\in \mathbb{Z}_{p}^{*}$. Then compute:
    $$
    K' = K^{-1/z},L' =  L^{1/z},\{K'_{x} = K_x^{1/z}\}_{x\in S}
    $$
    It sets the transform key TK and retrieve key RK as:
    $$
    TK = (S, K', L', \{K'_{x}\}_{x\in S})
    $$
    $$
    RK = z
    $$
    \item $Transform_{out}(TK, CT) \rightarrow CT'$ or $\perp$. The transform algorithm takes as input a transformation key $TK = (S, K', L', \{K'_{x}\}_{x\in S})$ and a ciphertext $CT = ((M,\rho), C, C', \{C_i, D_i\}_{i\in {[1,...,l]}})$. It first finds a set $W = \{\omega_i \in \mathbb{Z}_p\}_{i \in I}$ such that $\sum_{i\in I}\omega_i\lambda_i = s$, where $I$ is the set $I = \{i: \rho(i) \in S \}$ if S satisfies $(M,\rho)$, and outputs $\perp$ if S does not satisfy the access structure. The transformation algorithm computes:
    $$
    T = e(C', K') \cdot e(\prod_{i\in I}C_i^{\omega_i}, L') \cdot \prod_{i\in I}e(D_i^{w_i}, K'_{\rho_i})
    $$
    Finally, it sets the transform ciphertext $CT'$ as:
    $$
    CT' = (C, T)
    $$
    
    Note that the computation of $T$ is performed by an untrusted DCS and the set $W$ is found by DU, DCS may need to generate a zk proof to prove its correctness. We eliminate the division or inverse operations in the transform algorithm, making our scheme more zk-friendly.
    \item $Decrypt_{out}(CT', RK) \rightarrow \mathcal{M}$. The decrypt algorithm is executed by DO and takes as input a transformed(partially decrypted) ciphertext $CT' = (C, T)$ and a retrieve key $RK = z$, it computes:
    $$
    \mathcal{M} = C \cdot T^z
    $$
\end{itemize}

\subsection{Optional Single-Round Challenge Game using ZKP}
\label{subsec:challenge_game}
The results computed by DCS will be published on blockchain and were initially assumed to be correct and used by DO without delay based on optimistic assumption. Any challenger may deposit tokens to the smart contract to challenge the results within the challenge window. If an optional challenge happens, the challenger and the corresponding DCS engage in a game to determine who is correct, otherwise, the result will be finalized after the challenge window without challenge and the DCS does not need to generate proof. To reduce the challenge window, we use zk-SNARK to achieve a single-round challenge game. Once the computation result is challenged, the corresponding DCS must submit a zk-SNARK that can pass the smart contract verification to prove the correctness of the result within the response window, and it will be slashed otherwise. Part of the fine will be sent to the challenger as reward, while the other part will be compensated to DU. Due to different interests, challengers can be categorized into three distinct classes as follows.
\begin{itemize}
    \item DCS who submit the result: When a DCS challenges itself and wins as defender, it demonstrates the correctness of the on-chain result and retrieves the tokens staked for the challenge. However, it only wastes computational resources for generating the proof and gets nothing in this case. Conversely, if the DCS wins as challenger, some of the staked tokens will be forfeited to the DU, leading to a financial loss. Consequently, the DCS lacks the motivation to initiate a challenge.
    \item DU who outsourced the computation: The DU does not need to recompute the results; instead, they can detect the malicious behavior of the DCS through the subsequent decryption under the KEM setting. Upon a successful challenge, the DU will reclaim the payments made for the outsourced task and receive extra compensation. Therefore, even in the absence of other public challengers for verification, the DU is likely to initiate a challenge to protect their own interests.
    \item Other public challengers: Other challengers can leverage idle computing resources and publicly available data to recompute the results and identify any discrepancies. Upon discovering issues, they may stake tokens and initiate a challenge; a successful challenge will yield reward, while a failure will result in the loss of the staked tokens.
\end{itemize}

To achieve low and constant gas usage, we 
publish only the task data hash on blockchain while the original task data are stored on IPFS, and choose Plonk\cite{gabizon2019plonk} which has constant proof size and verification cost as proof system.

The indexed circuit satisfiability relation in our scheme is a set of tuples a set of triples $(\mathbbm{i}, \mathbbm{x}, \mathbbm{w})$ where $\mathbbm{x}$ is the public input: $hash(hash(CT||TK||\omega_i) || T)$, $\mathbbm{w}$ is the witness: $(CT, TK, \omega_i, T)$ and $\mathbbm{i}$ is the description of a circuit with constraints as follows:
\begin{enumerate}
    \item Constraint for partial decrypt result $T$: $$T == e(C', K') \cdot e(\prod_{i\in I}C_i^{\omega_i}, L') \cdot e(D_i^{w_i}, K'_{\rho_i})$$
    \item Constraint for public input: $$Public\_input == hash(hash(CT||TK||\omega_i)||T)$$
\end{enumerate}

We design a preprocessing zk-SNARK algorithm with specified circuit for our indexed circuit satisfiability relation based on Plonk proof system:
\begin{itemize}
    \item $Plonk.Preprocess(SRS, C) \rightarrow (pk,vk)$: The Preprocess algorithm takes as input a structured reference string(SRS) and a specified circuit $C$, outputs a proving key pk and a verifying key vk.
    \item $Plonk.Prove(pk,\mathbbm{x},\mathbbm{w}) \rightarrow \pi$: The Prove algorithm takes as input a proving key pk, a public input $hash(hash(CT||TK||\omega_i)||T)$ and a witness tuple $(CT, TK, \omega_i, T)$, outputs a proof $\pi$.
    \item $Plonk.Verify(vk,\mathbbm{x},\pi) \rightarrow 0/1$: The Verify algorithm takes as input a verifying key vk, a public input $hash(hash(CT||TK||\omega_i)||T)$ and a proof $\pi$, outputs 1 to indicate accept proof and 0 otherwise.
\end{itemize}

\subsection{Smart Contracts for Our Payable Scheme}
\label{subsec:smart_contract}
\begin{figure*}[tp]
    \centering
    \includegraphics[width=0.95\textwidth]{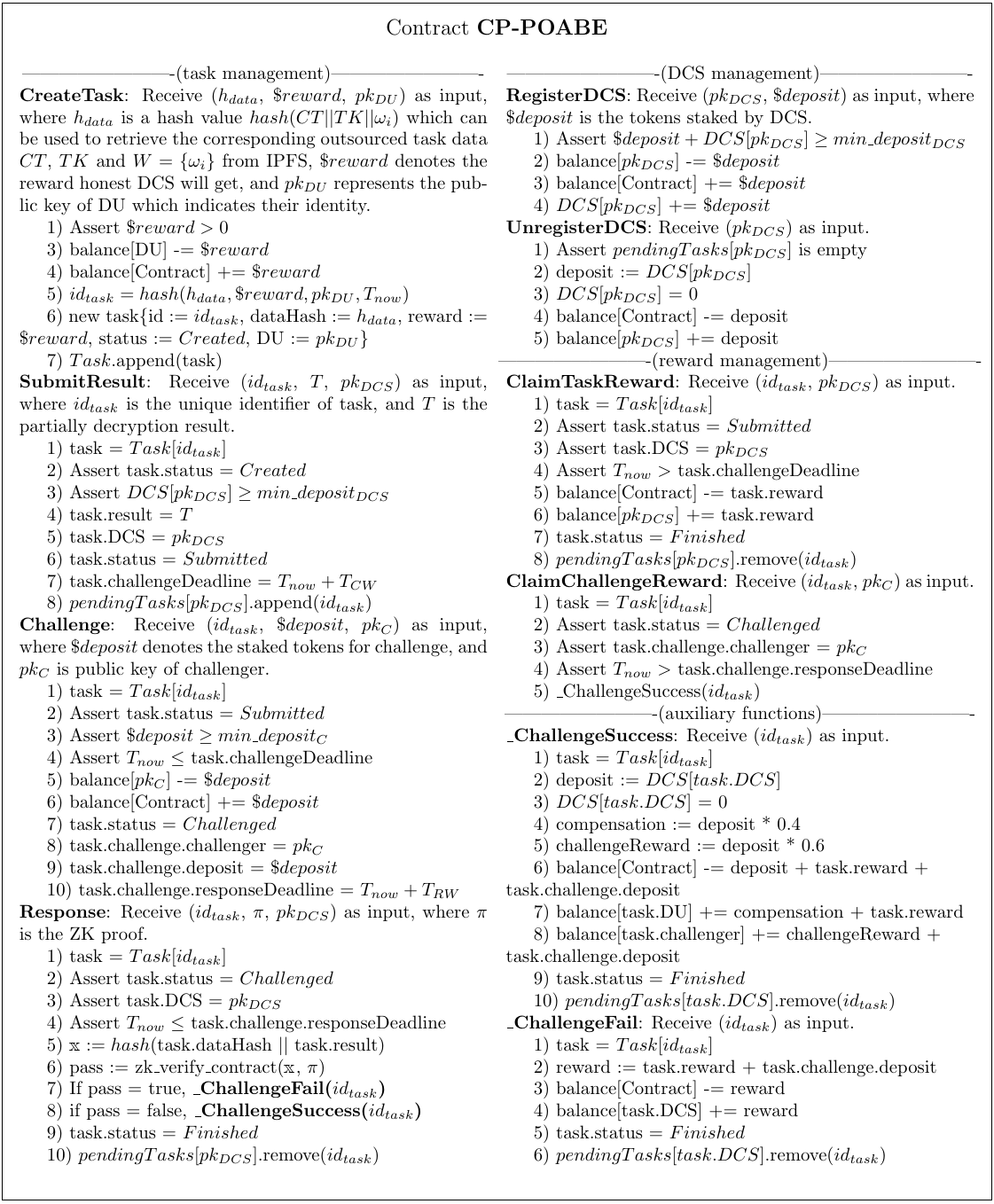}
    \caption{Contract design for CP-POABE.}
    \label{fig:opContract}
\end{figure*}

Our scheme contains two smart contracts: the CP-POABE contract and the ZKP verifier contract. The CP-POABE contract manages the whole process of outsourced decryption, and the ZKP verifier contract is invoked by the CP-POABE contract to verify the proof $\pi$ submitted by DCS.

\Cref{fig:opContract} presents the design of the CP-POABE contract. In the happy case of our scheme, DU firstly creates an outsourced decryption task by invoking \textbf{CreateTask}. Then any DCS could detect the new task, get corresponding outsourced task data from IPFS, compute the partially decrypted result, \textbf{SubmitResult} to blockchain, and get reward after challenge window. In the challenge case of our scheme, the challenger \textbf{Challenge} the result submitted by DCS. DCS should \textbf{Response} to the challenge by generating and submitting a ZK proof $\pi$ during the response window. The correctness of partially decrypted result is proven only if $\pi$ passes the verification of ZKP verifier contract.

To prevent malicious DCS from submitting incorrect results deliberately, a DCS must stake enough tokens(e.g., 5 ETH) through \textbf{RegisterDCS} before submitting results. A DCS that has no pending tasks can retrieve its deposit through \textbf{UnregisterDCS}; however, it will no longer be able to submit results to any task before it \textbf{RegisterDCS} again.

In the happy case, the DCS submitted the result can \textbf{ClaimTaskReward} after the challenge window. If a task result were challenged and DCS did not submit valid proof during the response window, the challenger could \textbf{ClaimChallengeReward}. \textbf{\_ChallengeSuccess} and \textbf{\_ChallengeFail} are auxiliary functions for disposing the deposit of DCSs and challengers.

In pseudo codes, $Task$ denotes the list of all tasks published in the blockchain, $pk_U$ denotes the public key of entity U, '\$' denotes tokens, $balance[U]$ denotes the balance of entity U, $T_{now}$ denotes current time, $T_{CW}$ denotes challenge window, $T_{RW}$ denotes response window, $min\_deposit_{C}$ denotes the minimal deposit for challenger, $min\_deposit_{DCS}$ denotes the minimal deposit for DCS, $DCS[U]$ denotes the amount of staked tokens by DCS U, and $pendingTasks[U]$ denotes the list of pending tasks of DCS U. Pending tasks refers to tasks that are \textbf{Submmited} or \textbf{Challenged}.

In the ZKP verifier contract, pairing check is necessary for verifying Plonk proof. Ethereum supports \textbf{BN256Pairing} precompile for pairing check and contains BLS-12381 precompiles in EIP-2537 in future upgrades. We will now use BN256 elliptic curve as our proof system base to facilitate the use of Ethereum BN256 precompiled contracts in verifier contract.

\begin{table*}[h!]
\centering
\caption{Comparison With Other Schemes}
\begin{tabular}{|c|c|c|c|c|c|c|}
\hline
Schemes & confidentiality & verifiability & exemptibility & fairness & decentralized outsourcing \\  %
\hline
\cite{lai2013attribute} & \Checkmark & \Checkmark & \XSolidBrush & \XSolidBrush & \XSolidBrush \\  %
\hline
\cite{li2013securely} & \Checkmark & \Checkmark & \XSolidBrush & \XSolidBrush & \XSolidBrush\\  %
\hline
\cite{qin2015attribute} & \Checkmark & \Checkmark & \XSolidBrush & \XSolidBrush & \XSolidBrush\\  %
\hline
\cite{lin2015revisiting} & \Checkmark & \Checkmark & \XSolidBrush & \XSolidBrush & \XSolidBrush\\  %
\hline
\cite{ma2015verifiable} & \Checkmark & \Checkmark & \XSolidBrush & \XSolidBrush & \XSolidBrush\\  %
\hline
\cite{wang2019fully} & \Checkmark & \Checkmark & \XSolidBrush & \Checkmark & \XSolidBrush\\  %
\hline
\cite{ge2023attribute} & \Checkmark & \Checkmark & \Checkmark & \Checkmark & \Checkmark\\  %
\hline
\cite{tao2024orr} & \Checkmark & \XSolidBrush & \XSolidBrush & \XSolidBrush & \Checkmark \\  %
\hline
\cite{mahdavi2024iot} & \Checkmark & \Checkmark & \XSolidBrush & \XSolidBrush & \XSolidBrush \\  %
\hline
\cite{hou2024blockchain} & \Checkmark & \Checkmark & \XSolidBrush & \Checkmark & \XSolidBrush \\  %
\hline
Our Scheme & \Checkmark & \Checkmark & \Checkmark & \Checkmark &\Checkmark \\  %
\hline
\end{tabular}
\label{table:functional comparison}
\end{table*}

\subsection{Security Analysis}
\textbf{Data Confidentiality.} The data confidentiality of our proposed CP-ABE with payable outsourced decryption scheme is based on the security of the CP-ABE with outsourced decryption scheme in \cite{green2011outsourcing} and \cite{ge2023attribute}. Note that the scheme proposed in \cite{ge2023attribute} is also based on the security of the scheme in \cite{green2011outsourcing}.
\begin{theorem}
    Suppose that the CP-OABE scheme in \cite{green2011outsourcing} is IND-CPA secure, then our proposed CP-ABE with payable outsourced decryption scheme is CPA secure in the KEM setting.
    \begin{proof}
        Suppose there exists a probabilistic polynomial-time(PPT) adversary $\mathcal{A}$ that can attack our scheme in the CPA security model for outsouring with advantage $\epsilon$, then we can build a simulator $\mathcal{B}$ that can break the CPA security of the CP-OABE scheme in \cite{green2011outsourcing}  with advantage $\epsilon$.
        
        \textit{Init.} The simulator $\mathcal{B}$ calls $\mathcal{A}$ to get the challenge access structure $(M^*,\rho^*)$. and sends it to the challenger $\mathcal{C}$ in the CP-OABE scheme in \cite{green2011outsourcing} as the structure on which it wishes to be challenged.

        \textit{Setup.} The simulator $\mathcal{B}$ gots public key $PK = (g, e(g,g)^{\alpha}, g^a, F)$ from the challenger $\mathcal{C}$ and returns it to the adversary $\mathcal{A}$.

        \textit{Phase 1.} The simulator $\mathcal{B}$ answers the adversary $\mathcal{A}$'s queries as follows:
        \begin{itemize}
            \item $Keygen(S)$ query: The simulator $\mathcal{B}$ passes it to the challenger $\mathcal{C}$ and returns the result that answered by $\mathcal{C}$ to $\mathcal{A}$.
        \end{itemize}

        \textit{Challenge.} $\mathcal{A}$ submits a encapsulated key pair $(K_0^*, K_1^*)\in \{0,1\}^{2\times k}$ to $\mathcal{B}$, and $\mathcal{B}$ sends them to the challenger $\mathcal{C}$. The challenger $\mathcal{C}$ generate the challenge ciphertext pair $(CT_0^*, CT_1^*)$ corresponds to $(K_0^*, K_1^*)$ and sends it to $\mathcal{B}$.

        \textit{Phase 2.} The simulator $\mathcal{B}$ continues to answer queries as in phase 1.

        \textit{Guess.} $\mathcal{B}$ passes ciphertext pair $(CT_0^*, CT_1^*)$ to $\mathcal{A}$ and gets the response bit 0 or 1. $\mathcal{B}$ outputs $\mathcal{A}$'s response as its guess.

        Note that, the simulation of $\mathcal{B}$ is perfect if the above game does not abort. The advantage of the simulator that wins the CPA game is the same as a adversary that breaks the CPA security of the CP-OABE in \cite{green2011outsourcing}. Thus, our proposed scheme is CPA secure.
    \end{proof}
\end{theorem}

\textbf{Verifiability.} On the one hand, the verifiability of our scheme is achieved through Zero-Knowledge Proof (ZKP), where the soundness of ZKP ensures the soundness of our verification. On the other hand, the verification algorithm is implemented via smart contract, and the verification will be executed on Ethereum. The soundness of this part is guaranteed by Ethereum’s consensus protocol, and the results will be validated and agreed upon by Ethereum nodes.

\textbf{Exemptibility.} We introduce a single-round challenge game through responsive ZKP and smart contract for DCS to prove the correctness of submitted results. The security is guaranteed by the completeness of ZKP and the soundness of verifiability analyzed above.

\textbf{Fairness.} Our payable scheme is based on Ethereum smart contract and token mechanism, where fairness is guaranteed by the public smart contract code and Ethereum consensus protocol. The fairness of our scheme is threefold. The DU must deposit tokens into the smart contract to submit the outsourced decryption task, and the smart contract will execute the payment either after the challenge window or upon verifying a valid proof. If the DCS submits the correct result, it will receive the reward at the end of the challenge window or after submitting a valid ZK proof, which is ensured by exemptibility. Otherwise, it will be slashed for malicious behavior. From the challenger's perspective, they will be rewarded for a valid challenge and slashed for a wrong challenge. Thus the proposed scheme achieves the fairness property as long as the exemptibility and Ethereum consensus protocol is guaranteed.

\begin{figure*}[h!]
    \centering
    \begin{subfigure}{0.3\textwidth}
        \centering
        \includegraphics[width=\linewidth]{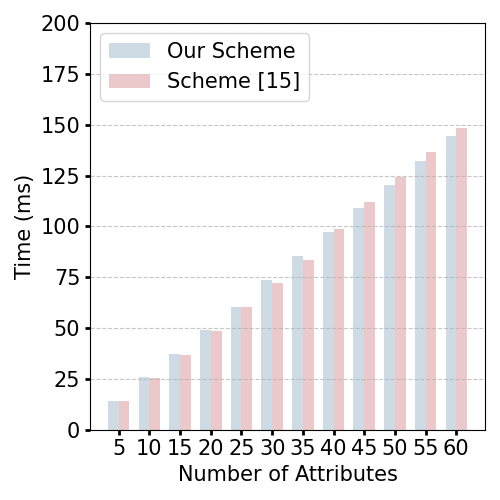}
        \subcaption{\textbf{Keygen}} 
        \label{fig:keygen_comparison}
    \end{subfigure}%
    \hfill
    \begin{subfigure}{0.3\textwidth}
        \centering
        \includegraphics[width=\linewidth]{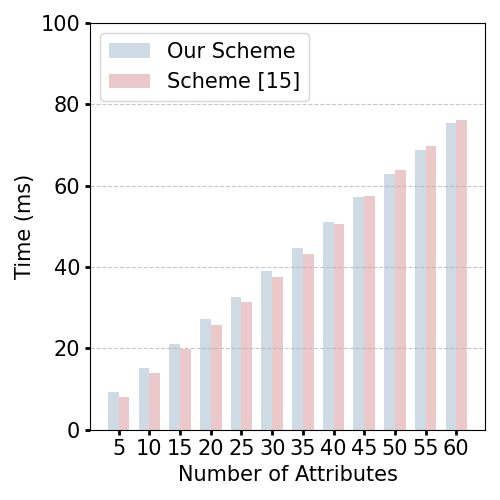}
        \subcaption{\textbf{Tkgen}}
        \label{fig:tkgen_comparison}
    \end{subfigure}%
    \hfill
    \begin{subfigure}{0.3\textwidth}
        \centering
        \includegraphics[width=\linewidth]{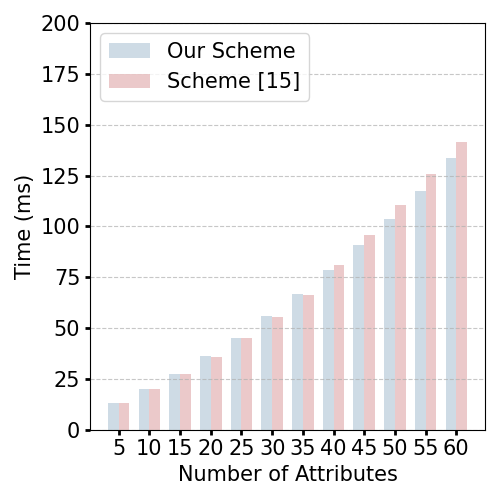}
        \subcaption{\textbf{Encrypt}}
        \label{fig:encrypt_comparison}
    \end{subfigure}
    \hfill
    \begin{subfigure}{0.3\textwidth}
        \centering
        \includegraphics[width=\linewidth]{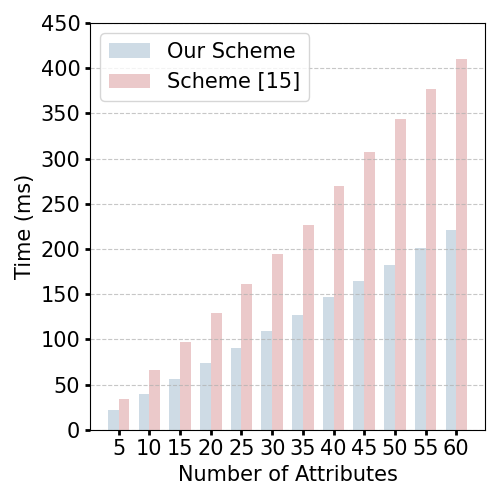}
        \subcaption{\textbf{Transform}}
        \label{fig:transform_comparison}
    \end{subfigure}%
    \hspace{10mm}
    \begin{subfigure}{0.3\textwidth}
        \centering
        \includegraphics[width=\linewidth]{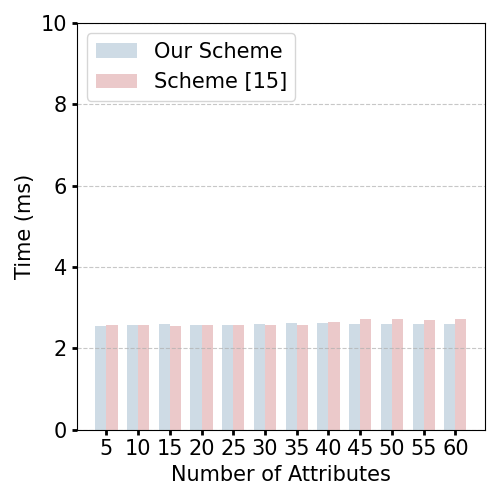}
        \subcaption{\textbf{Decrypt}}
        \label{fig:decrypt_comparison}
    \end{subfigure}%
    \caption{Comparison of (a) Keygen time, (b) Tkgen time, (c) Encrypt time, (d) Transform time and (e) Decrypt time for different numbers of attributes between \cite{ge2023attribute} and ours.}
    \label{fig:abe comparison}
\end{figure*}

\textbf{Self-challenge attack resistance.} As decribed in \ref{subsec:challenge_game}, DCS lacks motivation to self-challenge. As a defender, if it wins, it gains nothing but expends computation resources in generating proof. Conversely, if it wins as a challenger, a significant portion of the tokens previously staked will be slashed. Thus we achieve the property through tokenomics.

\begin{figure*}[h!]
    \centering
    \begin{subfigure}[t]{0.3\textwidth}
        \centering
        \includegraphics[width=\linewidth]{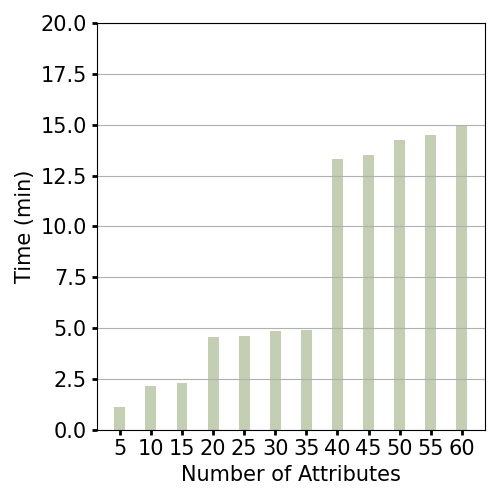}
        \subcaption{\textbf{Inner prove} \label{fig:inner_prove_time}}
    \end{subfigure}%
    \hfill
    \begin{subfigure}[t]{0.3\textwidth}
        \centering
        \includegraphics[width=\linewidth]{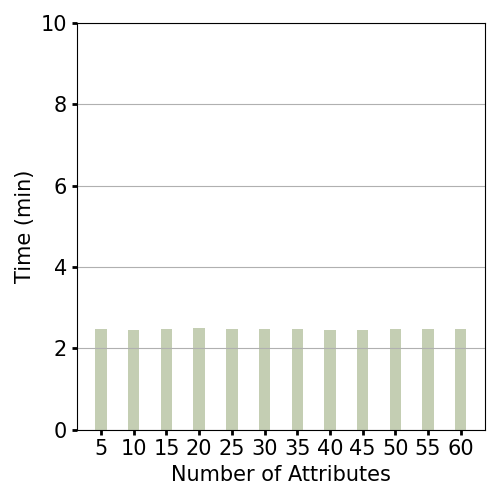}
        \subcaption{\textbf{Recursive prove}}
        \label{fig:recursive_prove_time}
    \end{subfigure}%
    \hfill
    \begin{subfigure}[t]{0.3\textwidth}
        \centering
        \includegraphics[width=\linewidth]{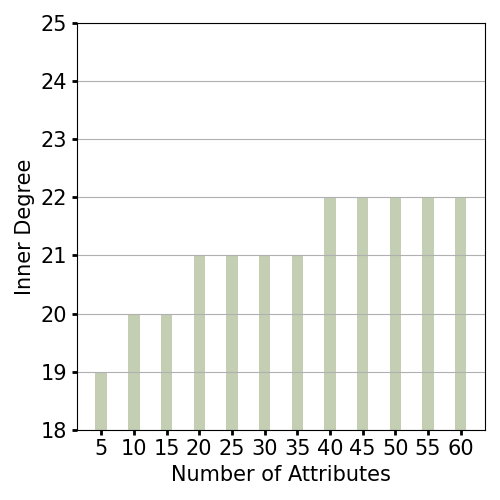}
        \subcaption{\textbf{Inner degree}}
        \label{fig:inner_degree}
    \end{subfigure}
    \caption{(a) Inner prove time, (b) Recursive prove time and (c) Inner circuit degree for different numbers of attribute.}
    \label{fig:zkp comparison}
\end{figure*}

\section{Performance And Evaluation}
\label{sec:Performance And Evaluation}
\subsection{Characteristic Comparison With Other Schemes}
We compare our scheme with the previous OABE schemes in terms of confidentiality, verifiability, exemptibility, fairness and decentralized outsourcing. \Cref{table:functional comparison} shows that almost all schemes achieve confidentiality and verifiability. However, only schemes \cite{ge2023attribute}, \cite{wang2019fully} and our scheme achieve fairness, and only \cite{ge2023attribute}, \cite{tao2024orr} and our scheme achieve decentralized outsourcing. Moreover, only our scheme and \cite{ge2023attribute} achieves exemptibility property. 
\renewcommand{\arraystretch}{1.5} %

\begin{figure}[tp]
    \centering
    \begin{minipage}{0.45\textwidth}
        \centering
        \includegraphics[width=\linewidth]{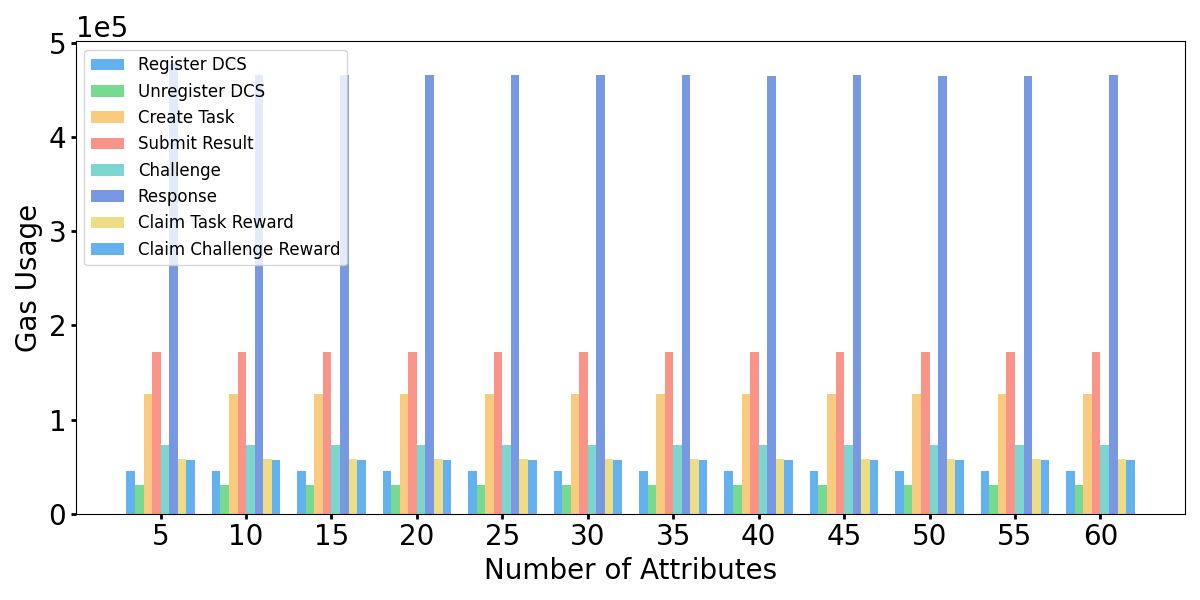}
        \subcaption{Gas usage of CP-POABE contract functions.}
        \label{fig:op_gas_usage_figure}
    \end{minipage}%
    \par\vspace{0.4cm}
    \begin{minipage}{0.45\textwidth}
        \centering
        \includegraphics[width=\linewidth]{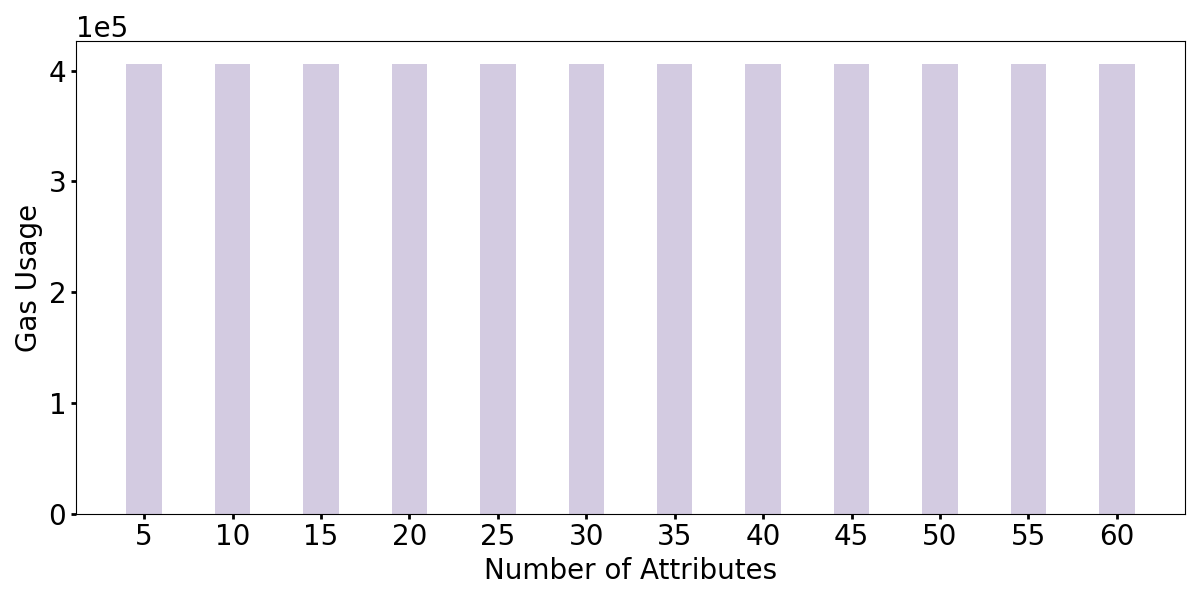}
        \subcaption{Gas usage of ZKP \textbf{Verify} function.}
        \label{fig:zkp_verify_gas_usage_figure}
    \end{minipage}
    \par\vspace{0.3cm}
    \begin{minipage}{0.45\textwidth}
        \centering
        \includegraphics[width=\linewidth]{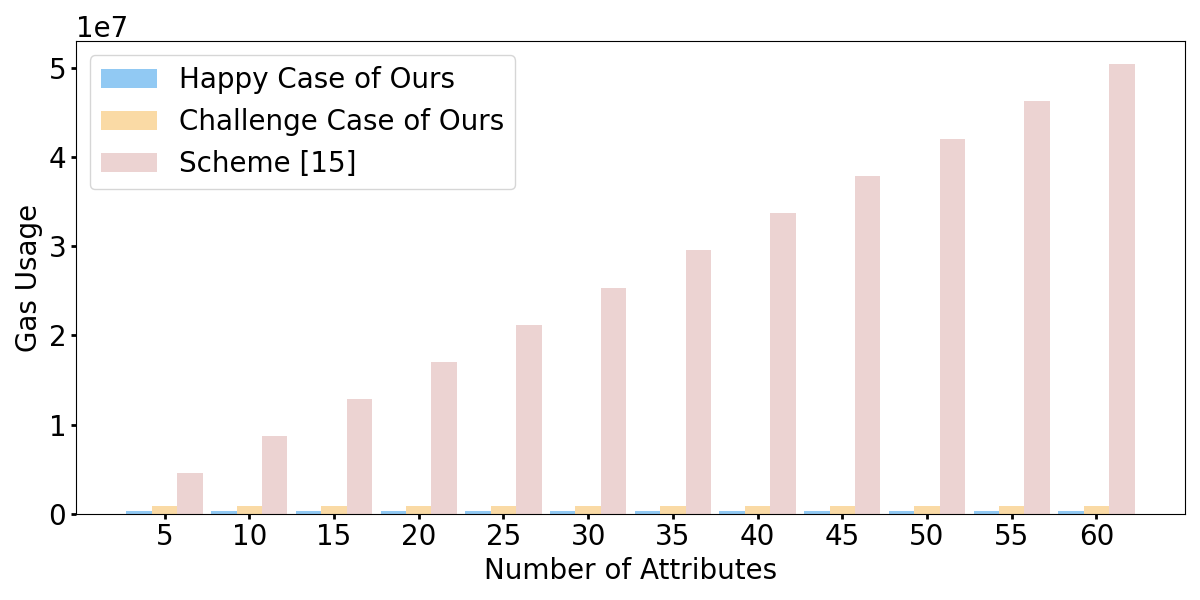}
        \subcaption{Gas usage comparison.}
        \label{fig:gas_usage_compare_figure}
    \end{minipage}
    \caption{Gas usage of (a) CP-POABE contract, (b) ZKP verifier contract and (c) Comparison among happy case of our scheme, challenge case of our scheme and scheme \cite{ge2023attribute}.}
    \label{fig:combined_gas_usage}
\end{figure}

\subsection{Experimental Evaluation}
We conducted experiment and performance analysis in three aspects: computation cost in ABE, zero-knowledge proof generation cost, and smart contract gas usage. In our experiment, the
access policy is set as an AND gate of attributes and its size varies from 5 to 60, with a step size of 5.
\subsubsection{ABE Computation Cost}
We implement our POABE scheme and OABE scheme in \cite{ge2023attribute} in Rust language with Barreto-Naehrig (BN) curve and SHA-256 hash function, and run benchmarks over a 2023 Apple MacBook Pro with M3 Max chip, 14 cores and 36GB RAM. For each algorithm, we repeat 1,000 times to get the average execution time.

As shown in \Cref{fig:abe comparison}, our scheme and the scheme \cite{ge2023attribute} exhibit a linear correlation in computation time for keygen, tkgen, encrypt, and transform relative to the number of attributes, while the decryption time remains approximately 2.6 ms for both. The time performance of our scheme is very close to \cite{ge2023attribute}, except in the case of the transform algorithm. For the transform algorithm, we only consider single-thread execution time on a local machine. As illustrated in \Cref{fig:transform_comparison}, our transform time is significantly lower than \cite{ge2023attribute}.

\subsubsection{Zero-Knowledge Proof Generation Cost}
We implemented the Halo2 circuit\cite{soureshjani2023automated} in Rust language by leveraging the open-source framework Scroll Halo2\footnote{https://github.com/scroll-tech/halo2}. Scroll Halo2 is based on Plonk\cite{gabizon2019plonk} polynomial interactive oracle proof(PIOP), Plonkish arithmetizations, and KZG\cite{kate2010constant} polynomial commitment scheme(PCS). 
Our circuit comprises an inner circuit which corresponds to indexed circuit satisfiability
relation described in \ref{subsec:challenge_game} and a recursive circuit. The inner circuit is used to prove the correctness of the partially decrypted(transform) result with outsourced task data and the recursive circuit compresses proof size through proving previous snark is valid. The final proof generation consists of two phases: in the first phase, we generate an inner proof using the inner circuit; in the second phase, we generate a final proof that can be verified by on-chain verifier contract using the recursive circuit. DEGREE is a variable specifies to set the circuit to have $2^{DEGREE}$ rows in Halo2. This variable determines the size of the instances that can be proven and significantly affects the proof generation time. In our experiments, the degree of the inner circuit ranged from 19 to 22, while the degree of the recursive circuit was fixed at 23.

We run benchmarks over a 2023 Apple MacBook Pro with M3 Max chip, 14 cores, and 36GB RAM, and repeat 10 times for each case to get the average time. The proof generation time and the degree for the inner snark are illustrated in \Cref{fig:inner_prove_time} and \Cref{fig:inner_degree}. The inner prove time increases with the degree linearly and slightly increases with the number of attributes when the degree is fixed. The recursive prove time is shown in \Cref{fig:recursive_prove_time} and approaches a constant since the degree of the recursive circuit is fixed at 23.

Due to the complexity of pairing and scalar multiplication operations, the proof generation time remains relatively long. However, our scheme does not rely on validity proof and utilizes an optimistic approach and challenge game instead. Consequently, proof generation will not become a bottleneck for the system. Moreover, since the proof generation takes on the level of minutes, we can reduce both the challenge window and the response window to the level of hours, rather than several days required by some optimistic rollups\cite{bousfield2022arbitrum}\cite{optimism}.

\subsubsection{Smart Contract Cost}
We implement our designed CP-POABE smart contract in Solidity which is described in Section \ref{subsec:smart_contract} and run the experiment over Lenovo Thinkbook14 G5+ IRH with Intel Core i7-13700H CPU and 32GB RAM. 
We set up a local Ethereum test network using Hardhat\footnote{https://hardhat.org/} and depoly the CP-POABE contract on it. We invoke each contract function 100 times to get their average gas usage, which is shown in \Cref{fig:op_gas_usage_figure}. We also conducted a separate test of the ZKP Verifier contract to evaluate the gas usage for ZKP verification in \Cref{fig:zkp_verify_gas_usage_figure}, which constitutes the most costly aspect of our CP-OPABE contract.

As \Cref{fig:op_gas_usage_figure} and \Cref{fig:zkp_verify_gas_usage_figure} illustrated, our contract functions gas usage are all almost constant while the gas usage of creating outsource task(contains upload CT and upload TK) and pairing operation computation in \cite{ge2023attribute} are increasing linearly as the number of attributes increase. \Cref{fig:gas_usage_compare_figure} shows the gas usage comparison among happy case of our scheme, challenge case of our scheme and scheme \cite{ge2023attribute}. Note that we can not test the gas usage of pairing operations of scheme \cite{ge2023attribute} in Ethereum because they are not compatible with native Ethereum, and thus we adopt the experimental result which is about 651304 for each meta pairing computation from \cite{ge2023attribute} for this part. The result shows that our gas usage is constant and is 11$\times$ to 140$\times$ in happy case and  4$\times$ to 55$\times$ in challenge case lower than \cite{ge2023attribute} in attribute numbers from 5 to 60.

\section{Conclusion}
\label{sec:Conclusion}
In this paper, we proposed a payable outsourced decryption attribute based encryption scheme based on Ethereum and then presented a formal security proof of its confidentiality. Our proposed scheme achieves verifiability, exemptibility, fairness and decentralize outsourcing without redundant information, and addresses heavy on-chain computation problem in \cite{ge2023attribute} through responsive zero-knowledge proof. Additional, we introduced a single-round challenge game under optimistic assumption to addresses the high computational cost of proof generation. Finally, we implemented and evaluated our scheme on Ethereum, and the results showed that our scheme is feasible, efficient, and has low and constant gas consumption even in the worst case.


\bibliography{reference}
\bibliographystyle{IEEEtran}

\end{document}